\documentclass[11pt]{article}

\usepackage[T1]{fontenc}
\usepackage[utf8]{inputenc}
\usepackage{lmodern}
\usepackage{amsmath,amssymb}
\usepackage{graphicx}
\usepackage{geometry}
\usepackage{microtype}
\usepackage{array}
\usepackage{hyperref}
\hypersetup{colorlinks=true,linkcolor=blue,citecolor=blue,urlcolor=blue}
\newcommand{\ManuscriptFigure}[1]{%
  \IfFileExists{#1}{\includegraphics[width=0.85\linewidth]{#1}}{%
    \fbox{\parbox[c][36mm][c]{0.79\linewidth}{\centering
    Figure file missing: \texttt{\detokenize{#1}}\\
    Supply this PDF for the final arXiv submission.}}}}

\title{An Accurate Zeroth-Order Perturbation Theory for Solving
Power-Law Potentials within the Framework of the Asymptotic Iteration Method}
\author{A. J. Sous\\
\small Faculty of Technology and Applied Sciences, Al-Quds Open University,\\
\small Nablus, Palestine
\and Ibsal Assi\\
\small Department of Physics and Physical Oceanography,\\
\small Memorial University of Newfoundland, St. John's, NL A1B 3X7, Canada
\and Nasser Saad\\
\small School of Mathematical and Computational Sciences,\\
\small University of Prince Edward Island, Charlottetown, PE C1A 4P3, Canada}
\date{}

\begin{document}
\maketitle

\begin{abstract}
Spherically symmetric singular potentials have a wide range of applications, such as molecular and atomic physics. In this work, we present a zeroth-order perturbation theory for solving the Schr\"odinger equation within the framework of the asymptotic iteration method (AIM). We rely on exponential and hyperbolic potential ansatz substitutions, where the difference between the original and new potential is treated as a small perturbation.
Each potential ansatz has a free parameter chosen small enough to give an accurate zeroth-order perturbative wave equation solution.
This method avoids the oscillatory behavior observed in the asymptotic iteration solutions when applying the AIM to such problems. Finally, we illustrate our approach by solving a few problems, particularly potentials with quartic and sextic singularities. We compare our results with those obtained via other methods and find good agreement.
\end{abstract}


\section{Introduction}
\label{intro}
Exact and numerical solutions of the Schr\"odinger equation are of both theoretical and experimental importance. Hamiltonians with spherically symmetric potentials, in particular, have found a wide range of applications in molecular and nuclear physics \cite{d2000,a2002,j2019}. For example, Lennard-Jones' potential (aka six-twelve potential) is used to model interactions in neutral molecules \cite{s2013}. It is well-known that the space of exactly solvable solutions of the Schrodinger equation is relatively limited; hence there is always a need for numerical techniques to solve the problem under consideration. The Lagrange-mesh methods \cite{d2015}, methods of finite difference \cite{HOFD,FD2}, and the asymptotic iteration method (AIM) \cite{AIM1}  are some of the numerical methods attributed to solving many eigenvalue problems.
\vskip0.1true in
\noindent In this work, we use AIM to study the power-law potentials with spherical symmetry. In practice, the application of AIM to such problems is limited due to the oscillatory behaviour in the numerical computations generated by AIM iterative process; therefore, convergence is not easily obtained in several cases \cite{AIM_SP1,AIM_SP2,AIM_SP3}. 
\vskip0.1true in
\noindent An earlier proposal to overcome this problem was introduced in \cite{AIM_PT}, where a perturbative approach is proposed and applied cautiously. 
A reason for this careful examination has to do with the computational demand of the perturbative approaches and the need for higher-order perturbative calculations. 
The approach was also highly sensitive to the asymptotic wave-function solution required to put the eigenvalue problem in the appropriate form for AIM. Furthermore, the convergence rate was highly sensitive to the initial seed point $x_0$ needed to utilize the iterative procedure.
\vskip0.1true in
\noindent In the present work, we introduced, in the next section, a less demanding approach where we formulate the problem in such a way that we obtain the results we are looking for within a reasonable accuracy and manageable computational time. In Section \ref{sec:PA}, we initiate this technique using two different classes of potential ansatz. In Section 4, we provide applications to our method by solving three classes of problems, and finally, we conclude our work in Section 5.

\section{Zeroth-order perturbation theory within the frame of the AIM}
\label{sec:PT_AIM}
The radial Schrodinger equation under consideration is
\begin{equation}
    \left[-\frac{1}{2}\frac{d^2}{dr^2}+V(r)\right]\psi(r)=E\psi(r)
\end{equation}
where $V(r)$ is the effective potential that is difficult to address using the straightforward application of AIM. Our starting point is to add and subtract an Ansatz potential $\mathcal{V}_0(\lambda,r)$ to the above equation, giving
\begin{equation}
    \left[-\frac{1}{2}\frac{d^2}{dr^2}+\mathcal{V}_0(\lambda,r)+U_{\lambda}(r)\right]\psi(r)=E\psi(r)
\end{equation}
where
\begin{equation}
    U_{\lambda}(r)=V(r)-\mathcal{V}_0(\lambda,r)
\end{equation}
is a regular function at the origin ($r=0$) and $\lambda$ that is small enough so that $U_\lambda(r)$ can be treated as a perturbation that satisfy  $\lim_{\lambda\to0}U_{\lambda}(r)=0$.  AIM is then used to solve the Schr\"odinger equation  
\begin{equation}\label{wf}
    \left[-\frac{1}{2}\frac{d^2}{dr^2}+\mathcal{V}_0(\lambda,r)\right]\psi^{(0)}_{\lambda}(r)=E^{(0)}_{\lambda}\psi^{(0)}_{\lambda}(r)
\end{equation}
Usually a coordinate transformation $y=y(r)$ and a wavefunction Ansatz $\psi^{(0)}_{\lambda}(y)=g(y)f(y)$ is used to map the above equation  into a form suitable to AIM application
\begin{equation}
\label{eqn:AIMeq}
    \frac{{{d}^{2}}f(y)}{d{{y}^{2}}}={{k}_{0}}(y,E^{(0)}_{\lambda})\frac{df(y)}{dy}+{{s}_{0}}(y,E^{(0)}_{\lambda})f(y)
\end{equation}
where
\begin{equation}
\label{eqn:k0}
    {{k}_{0}}(y,E^{(0)}_{\lambda})=-2\left( \frac{1}{g(y)}\frac{dg(y)}{dy}+\frac{y''}{2y{{'}^{2}}} \right)
\end{equation}
and
\begin{equation}
\label{eqn:s0}
    {{s}_{0}}(y,E^{(0)}_{\lambda})=-\frac{y{{'}^{2}}}{y{{'}^{2}}g(y)}\frac{{{d}^{2}g(y)}}{d{{y}^{2}}}-\frac{y''}{y{{'}^{2}}g(y)}\frac{dg(y)}{dy}+\frac{2\,{{\mathcal{V}}_{0}}(\lambda,y)-2E^{(0)}}{y{{'}^{2}}}.
\end{equation}
The AIM sequences  $k_n(y)$ and $s_n(y)$, $n = 1, 2,\cdots$ are generated recursively via
\begin{equation}
\label{eqn:AIMiter}
    \left(\begin{array}{cc}
    k_{n}(y)\\
    s_{n}(y)\\ 
    \end{array}\right)=\left(\begin{array}{cc}
    \frac{d}{dy}+k_{0} & 1\\
    s_{0}&\frac{d}{dy}\\
    \end{array}\right)\left(\begin{array}{cc}
    k_{n-1}(y)\\
    s_{n-1}(y)\\
    \end{array}\right).
\end{equation}
Assume that the ratio $s_{n}(y)/k_{n}(y)$ converges asymptotically to some function $\eta (y)$ for sufficiently large $n$ \cite{AIM1}, the energy spectrum are calculated as the roots of  the termination condition
\begin{equation}
\label{eqn:termination}
    \Delta_{n}(y,E^{(0)})=k_{n}(y)s_{n-1}(y)-k_{n-1}(y)s_{n}(y)
\end{equation}
and  the wave functions $\psi_\lambda^{(0)}(y) $, see \eqref{wf},  are establish using \cite{AIM1}
\begin{equation}
    f(y)={\rm exp}\left(-\int_{0}^{y}\eta(t)\, dt\right).
\end{equation}
In comparison with the perturbation theory in quantum mechanics \cite{FF},  $E^{(0)}$ is the zeroth-order perturbative solution of the original wave equation. We choose our perturbation parameter $\lambda$ such that the lowest-order solution gives accurate results; this is possible by controlling the free parameter $\lambda$ that comes with the potential  $\mathcal{V}_0(r)$. Higher-order corrections to the energy and wave function can be calculated via the well-known time-independent perturbation theory if needed. We may note the following two identities 
\begin{equation}
\label{eqn:WF_ex}
    \lim_{\lambda\to 0} \psi^{(0)}_{\lambda}(r)=\psi(r)
\end{equation}
and 
\begin{equation}
\label{eqn:en_ex}
    \lim_{\lambda\to 0} E^{(0)}_{\lambda}=E.
\end{equation}
%%%This block is excluded
\iffalse
According to time-independent perturbation theory [x], the first order correction to the $j^{\rm th}$ energy is
\begin{equation}
    E_j^{(1)}=\left\langle \psi_j^{(0)} \middle| U(r)\middle| \psi_j^{(0)} \right\rangle=\frac{\int_{y(0)}^{y(\infty)}f_j(y)U(y)f_j(y)g(y)g(y)\frac{dy}{y'}}{\int_{y(0)}^{y(\infty)}f_j(y)f_j(y)g(y)g(y)\frac{dy}{y'}}
\end{equation}
with the corresponding first-order correction to the wavefunction is
\begin{equation}
    \ket{\psi_n^{(1)}}=\sum_{m\neq n}\frac{\left\langle \psi_m^{(0)} \middle| U(r) \middle| \psi_n^{(0)} \right\rangle}{E^{(0)}_n-E^{(0)}_m}\ket{\psi_m^{(0)}}
\end{equation}
On the other hand, the second-order correction to the energy is
\begin{equation}
    E_n^{(2)}=\sum_{m\neq n}\frac{\left|\left\langle \psi_m^{(0)} \middle| U(r) \middle| \psi_n^{(0)} \right\rangle\right|^2}{E^{(0)}_n-E^{(0)}_m}
\end{equation}

{\color{red} Note: To evaluate the integrals, one might use some Gauss quadrature methods etc, where the endpoints are excluded.}
\fi

\section{Potential ansatz substitutions} 
\label{sec:PA}
Consider the following general form of the power-law potential
\begin{equation}\label{eq13}
    V(r)=\frac{a_1}{r}+\frac{a_2}{r^2}+\cdots+\frac{a_n}{r^n},
\end{equation}
we propose the following two Ansatz for $\mathcal{V}_0(\lambda,r)$
\begin{equation}
\mathcal{V}_0(\lambda,r)=\sum_{k=0}^{n}\mathcal{A}_k\left(e^{\lambda r}-1\right)^{-k} 
\label{eqn:EA}
\end{equation}
and
\begin{equation}
\mathcal{V}_0(\lambda,r)=\sum_{k=0}^{n}\mathcal{B}_k\left(\sinh{\lambda r}\right)^{-k} 
\label{eqn:HA}
\end{equation}
where the coefficients $\{\mathcal{A}_k,\mathcal{B}_k\}$ are chosen such that
\begin{equation}
    U_{\lambda}(r)=V(r)-\mathcal{V}_0(\lambda,r)\to 0, \quad as\quad \lambda\to 0.
\end{equation}
Generally speaking, $\lambda$ is a free parameter chosen to be small such that $U_\lambda(r)$ is negligible compared to $\mathcal{V}_0(\lambda,r)$. We refer to the Ansatz in Eq.\eqref{eqn:EA} by the `\emph{Exponential Ansatz}' (EA), while the Ansatz in Eq.\eqref{eqn:HA} is referred to as the `\emph{Hyperbolic Ansatz} (HA). 
\vskip0.1true in
\noindent In Appendix \ref{App:Ansatz}, we have listed the Ansatz choices \eqref{eqn:EA}-\eqref{eqn:HA} for power-low potentials with up to sextic singularities. In the next section, we apply those Ansatz to a variety of problems and find the corresponding discrete spectra using the AIM.   

%For our case, we will use a change of variable of the form $y=1-2e^{-\lambda r}$ and a wavefunction Ansatz 
%\begin{equation}
%    \psi^{(0)}(y)=\mathcal{N}g(y)f(y)=\mathcal{N}g(y){\rm exp}\left(-\int_{0}^{y}dt\frac{s_n(t)}{k_n(t)}\right)
%\end{equation}
%where $g(y)=(1-y)^\alpha(1+y)^\beta$ for some positive parameters $\{\alpha,\beta\}$ and $\mathcal{N}$ is the normalization constant.

\section{Applications}
\subsection{The Coulomb potential}
\begin{figure}[ht]
    \centering
    \ManuscriptFigure{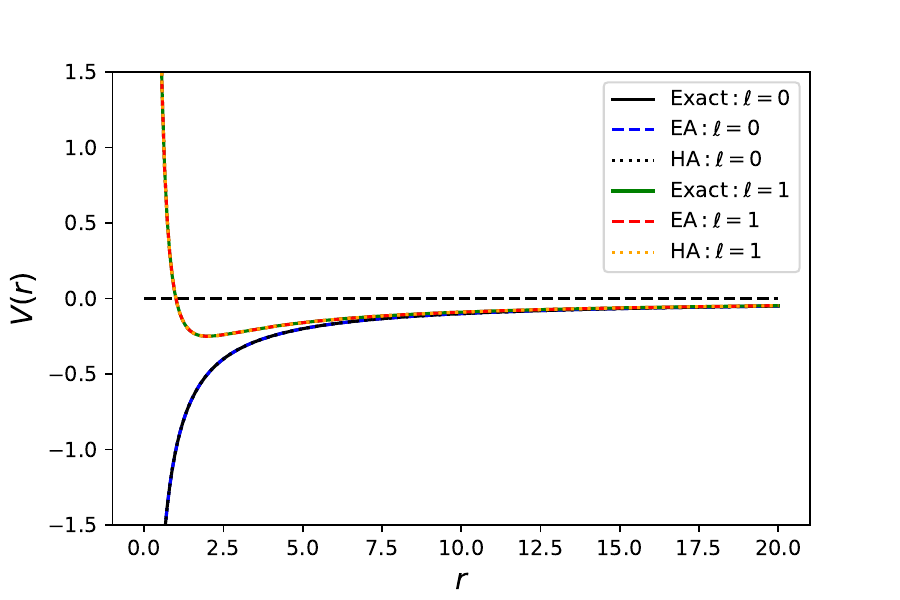}
    \caption{A plot of Coulomb potential and the Ansatz potentials for $Q=1$ and $\ell={0,1}$.}
    \label{fig:col}
\end{figure}

One of the simplest and exactly solvable potentials is the Coulomb problem,
\begin{equation}
	    V(r)=-\frac{Q}{r}+\frac{\ell(\ell+1)}{2r^2}
\end{equation}
where $Q$ is the charge of the particle (e.g. electron) and $\ell=0,1,\dots$ is the angular momentum quantum number. The above potential corresponds to the following exact energy spectrum formula 

\begin{equation}
\label{eqn:col_ex}
	    E_{n\ell}=-\frac{1}{2}\left(\frac{Q}{n+\ell+1}\right)^2,\qquad n=0,1,2,\cdots.
\end{equation}
Starting with the exponential Ansatz, we use the substitutions from Appendix \ref{App:Ansatz} for $v_1=1/r$, and $v_2=1/r^2$.  For the EA ansatz, use the change of variable 
$y=1-2e^{-\lambda r}$ and we take $g(y)=1-y^2$. For the resulting equation, Table \eqref{tab:CP_EA} reports the numerical eigenvalues obtained using AIM. For these values, the seed (initial) value $y_0=0$ was used, and the perturbation parameter $\lambda=0.001$. Table \eqref{tab:CP_EA} reflects the good agreement with the exact eigenvalues obtained using \eqref{eqn:col_ex}.  
\vskip0.1true in
\noindent We may also employ the HA to find the energy spectrum for the same parameters chosen above, and we compared the results with those obtained via the EA ansatz as shown in Table \ref{tab:CP_EAvsHA}. In our AIM calculations for the HA case, we applied the transformation $y=e^{-\lambda r}$ and $g(y)=1$, and we used the seed point $y_0=0.5$ with $\lambda=0.001$. We found that the two Ansatz are in good agreement with each other. Furthermore, we made a plot of the potential Ansatz for different values of the potential parameters and compared that with the exact potential as shown in Fig. \ref{fig:col}. 

%\begin{table}[h!]
%  \begin{center}
%    \caption{The lowest 8 bound state energies of Coulomb potential obtained by the zeroth order perturbation theory with the exponential ansatz for $Q=1$ and $\ell=0,1$ with several choices of $\lambda$.} 
%    \label{tab:EA_CP_exact}
%    \begin{tabular}{|c|c|c|c|c|c|} 
%      \hline
%      \multicolumn{3}{|c|}{$\ell=0$} & \multicolumn{3}{c|}{$\ell=1$}\\ % <-- Combining two cells with alignment c| and content 12.
%      \hline
%      $\lambda=0$ & $\lambda=0.001$ & $\lambda=0.0001$ &$\lambda=0$ & $\lambda=0.001$ & $\lambda=0.0001$\\
%      \hline
%     -0.5 &-0.50100012500000&-0.50010000125000&-0.125&-0.12575004166667&-0.12507500041667\\
%     -0.125&-0.12600050000000&-0.12510000500000&-0.0555555556&-0.05644504166667&-0.05564445041667\\
%    -0.0555555556&-0.05655668055556&-0.05565556680556&-0.03125&-0.03218894791667&-0.03134376447917\\
%    -0.03125&-0.03225200000000&-0.03135002000000&-0.02&-0.02096256166667&-0.02009602561667\\ 
%    -0.02&-0.02100312500000&-0.02010003125000&-0.0138888889&-0.01486504166667&-0.01398615041667\\
%   -0.0138888889&-0.01489338888889&-0.01398893388889&-0.0102040816&-0.01118922534014&-0.01030209633503\\
%-0.0102040816&-0.01121020663265&-0.01030414288265&-0.0078125&-0.00880429947917&-0.00791101174479\\
%-0.0078125&-0.00882050000000&-0.00791258000000&-0.0061728395&-0.00717004166667&-0.00627170041667\\ 
%        \hline
%    \end{tabular}
%  \end{center}
%\end{table}

\begin{table}[h!]
  \begin{center}
    \caption{The energy spectrum of the Coulomb potential obtained via the exact well-known formula versus the results obtained via the Exponential Ansatz with $Q=1$, $y_0=0$, $\lambda=0.001$, and 400 iterations.} 
    \label{tab:CP_EA}
\resizebox{\linewidth}{!}{%
    \begin{tabular}{|c|c|c|c|c|c|} 
      \hline
      \multicolumn{2}{|c|}{$\ell=0$} & \multicolumn{2}{c|}{$\ell=1$}& \multicolumn{2}{c|}{$\ell=2$}\\ % <-- Combining two cells with alignment c| and content 12.
      \hline
      \textbf{Exact} & \textbf{AIM} &\textbf{Exact} & \textbf{AIM}&\textbf{Exact} & \textbf{AIM}\\
      \hline
     -0.5 &-0.5000000419&-0.125&-0.1250004167  &-0.0555555556&-0.05555643056 \\ 
     -0.125&-0.1250005000&-0.0555555556&-0.05555659722  &-0.03125&-0.031251750000\\
    -0.0555555556&-0.05555668056&-0.03125&-0.03125191667  &-0.02&-0.02000287500 \\ 
    -0.03125&-0.03125200000&-0.02&-0.02000304167 &-0.0138888889&-0.01389313889 \\ 
    -0.02&-0.02000312500 &-0.0138888889&-0.01389330556  &-0.0102040816&-0.01020995663 \\ 
   -0.0138888889&-0.01389338889&-0.0102040816&-0.01021012330  &-0.0078125&-0.007820250000 \\ 
   
-0.0102040816&-0.01021020663&-0.0078125&-0.007820416667  &-0.0061728395&-0.006182714506 \\
-0.0078125&-0.007820500000&-0.0061728395&-0.006182881173 &-0.005&-0.005012250000 \\ 
        \hline
    \end{tabular}
}%
  \end{center}
\end{table}

\begin{table}[h!]
  \begin{center}
    \caption{$Q=1$ Hyperbolic Ansatz (HA) vs Exponential Ansatz (EA). In EA, we used $x_0=0$, $\lambda=0.001$, while in the HA, we took $x_0=0.5$, $\lambda=0.001$}
    \label{tab:CP_EAvsHA}
\resizebox{\linewidth}{!}{%
    \begin{tabular}{|c|c|c|c|c|c|c|c|c|} 
      \hline
      \multicolumn{2}{|c|}{$\ell=0$} & \multicolumn{2}{c|}{$\ell=1$}& \multicolumn{2}{c|}{$\ell=2$}\\ % <-- Combining two cells with alignment c| and content 12.
      \hline
      \textbf{EA} & \textbf{HA} & \textbf{EA} & \textbf{HA}& \textbf{EA} & \textbf{HA}\\
      \hline
      -0.5000000419&-0.4735505486&-0.1250004167 & -0.1249991667 &-0.05555643056& - 0.05555380558 \\
-0.1250005000&-0.1249990000&-0.05555659722& - 0.05555347228&-0.03125175000&- 0.03124650025 \\ 
-0.05555668056 &-0.05555330564&-0.03125191667&- 0.03124616706 &-0.02000287500&- 0.01999425121 \\
-0.03125200000 &-0.03124600047 &-0.02000304167& - 0.01999391823 &-0.01389313889&- 0.01388039293 \\
-0.02000312500 &-0.01999375176 &-0.01389330556 &- 0.01388006035& -0.01020995663&- 0.01019234249 \\
-0.01389338889 &-0.01387989409 &-0.01021012330 &- 0.01019201059 & -0.007820250000&- 0.007797025166 \\
-0.01021020663 &-0.01019184466&-0.007820416667& - 0.007796694296 &-0.006182714506&- 0.006153141844 \\
-0.007820500000 &-0.007796528896&-0.006182881173 &- 0.006152812476&-0.005012250000 &- 0.004975600141\\
        \hline
    \end{tabular}
}%
  \end{center}
\end{table}
 
\subsection{The spiked isotropic oscillator}
\begin{figure}[ht]
    \centering
    \ManuscriptFigure{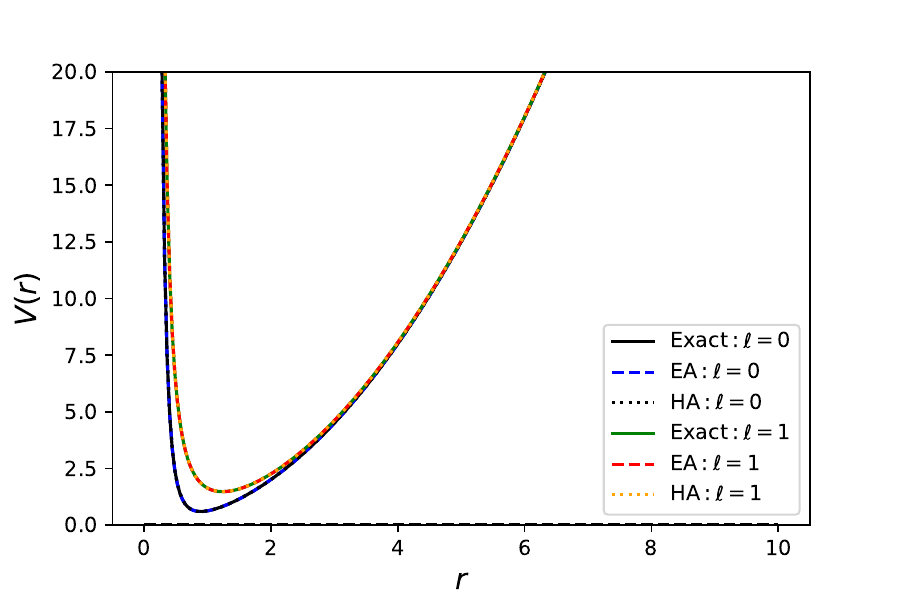}
    \caption{A plot of spiked isotropic oscillator potential and its Ansatz potentials for $\omega=1$, $a=0.5$ and $\ell={0,1}$.}
    \label{fig:spiked}
\end{figure}

Another interesting problem to apply our Ansatz method is the spiked oscillator with quartic singularity \cite{Spiked1,Spiked2,Spiked3,Spiked4,Spiked5,Spiked6,Spiked7}
\begin{equation}
\label{eqn:spiked}
	    V(r)=\frac{1}{2}\omega^2r^2+\frac{a^2}{2r^4}+\frac{\ell(\ell+1)}{2r^2}
\end{equation}
The potential Ansatz is chosen based on the transformations in Appendix \ref{App:Ansatz}. However, for the quadratic term, we use the identity $r^2={1}/({1}/{r^2})$, and then we use the corresponding substitution for ${1}/{r^2}$ from the appendix. For the hyperbolic ansatz, we took $y=\sinh(\lambda r)$, $g(y)=e^{-y^2/2}$, $y_0=1/10$ and $\lambda=0.01$. We took the parameters $\omega=1$ and $a=0.5$ and compared our results for the lowest ten energies obtained via the TRA \cite{Spiked7} as shown in Table \ref{tab:spiked_HA}. We find good agreement between our results and those obtained via the TRA. For the EA, we took $y=e^{-\lambda r}-1$, $g(y)=1$, $y_0=1/15$ and $\lambda=0.01$. A comparison between the results of both ansatzes is given in Table \ref{tab:spiked_EAvsHA}. We also plotted our Ansatz and compared them with the exact potential \eqref{eqn:spiked} as shown in Fig. \eqref{fig:spiked}. 

%We consider the following potential Ansatz
%\begin{equation}
%    {{\mathcal{V}}_{0}}(r)={{\lambda }^{2}}\left( \frac{\ell \left( \ell +1 \right)}{2}+\frac{{{a}^{2}}{{\lambda }^{2}}}{3} \right)\frac{1}{{{\sinh }^{2}}\left( \lambda r \right)}+\frac{{{a}^{2}}}{2}\frac{{{\lambda }^{4}}}{{{\sinh }^{4}}\left( \lambda r \right)}+\frac{1}{2}\frac{{{\omega }^{2}}}{{{\lambda }^{2}}}{{\sinh }^{2}}\left( \lambda r \right)+\frac{\ell \left( \ell +1 \right)}{2}\frac{{{\lambda }^{2}}}{3}-\frac{{{a}^{2}}}{2}\frac{{{\lambda }^{4}}}{45}
%\end{equation}

%And we solve the following unperturbed problem
%\begin{equation}
%    \left[ -\frac{1}{2}\frac{{{d}^{2}}}{d{{r}^{2}}}+{{\mathcal{V}}_{0}}(r) \right]{{\psi }^{\left( 0 \right)}}(r)={{E}^{\left( 0 \right)}}{{\psi }^{\left( 0 \right)}}(r)
%\end{equation}
%Starting with the HA $\mathcal{V}_{0}(r)$, we perform a change of variable $y=\sinh(\lambda r)$, giving 
%\begin{equation}
%    \frac{{{d}^{2}}{{\psi }^{\left( 0 \right)}}(y)}{d{{y}^{2}}}=-\frac{y}{\left( {{y}^{2}}-1 \right)}\frac{d{{\psi }^{\left( 0 \right)}}(y)}{dy}+\frac{2\left( {{\mathcal{V}}_{0}}(y)-{{E}^{\left( 0 \right)}} \right)}{{{\lambda }^{2}}\left( {{y}^{2}}-1 \right)}{{\psi }^{\left( 0 \right)}}(y)
%\end{equation}
%which is the AIM form. In our calculations we took $a=0.5$, $\omega=1$, $y_0=0.1$ and $\lambda=0.01$ , where the results are summarized in Table \ref{tab:spiked_HA}.  

\begin{table}[h!]
  \begin{center}
    \caption{The first few lowest energies of the spiked oscillator Eq. \ref{eqn:spiked} for $\omega=1$, $a=0.5$ obtained via the TRA \cite{Spiked7} versus the Hyperbolic Ansatz (HA). For the HA, we took $y_0=0.1$, $\lambda=0.01$ and 400 iterations.}
    \label{tab:spiked_HA}
\resizebox{\linewidth}{!}{%
   \begin{tabular}{|c|c|c|c|c|c|c|c|} 
      \hline
      \multicolumn{2}{|c|}{$\ell=3$} & \multicolumn{2}{c|}{$\ell=4$}& \multicolumn{2}{c|}{$\ell=5$}&\multicolumn{2}{c|}{$\ell=6$}\\ % <-- Combining two cells with alignment c| and content 12.
      \hline
      \textbf{TRA} & \textbf{AIM} & \textbf{TRA} & \textbf{AIM}&\textbf{TRA} & \textbf{AIM}&\textbf{TRA} & \textbf{AIM}\\
      \hline
      4.514093664&4.514506437&5.507895881&5.508494706&6.505038139&6.505852852&7.503491750&7.504555983\\
      6.520199526&6.521143478&7.510734904&7.511986392&8.506579901&8.508146175&9.504420052&9.506335250\\
8.526201793&8.527889392&9.513558855&9.515663690&10.50811814&10.51063631&11.50534730&11.50831359\\
10.53215395&10.53474888&11.51636323&11.51952682&12.50965288&12.51332324&13.50627349&13.51049099\\
12.53804737&12.54172614&13.51916096&13.52357601&14.51118415&14.51620696&15.50719864&15.51286741\\
14.54389207&14.54882496&15.52194082&15.52781144&16.51271197&16.51928747&17.50812273&17.51544282\\
16.54955292&16.55604878&17.52470775&17.53223334&18.51423637&18.52256476&19.50904579&19.51821721\\
18.55529122&18.56340074&19.52745910&19.53684187&20.51575737&20.52603882&21.50996781&21.52119054\\
20.56054355&20.57088374&21.53019782&21.54163723&22.51727499&22.52970964&23.51088879&23.52436279\\
22.56580545&22.57850044&23.53292075&23.54661959&24.51878925&24.53357721&25.51180874&25.52773392\\
       % \hline
      %\multicolumn{2}{|c|}{$\ell=6$}&
      %\multicolumn{2}{c|}{$\ell=7$}&
      %\multicolumn{2}{c|}{$\ell=5$}\\
      %\hline
      %\textbf{TRA} & \textbf{AIM}&\textbf{TRA} & \textbf{AIM}\\%&\textbf{TRA} & \textbf{AIM}\\
      %\hline

      \hline
    \end{tabular}
}%
  \end{center}
\end{table}

\begin{table}[h!]
  \begin{center}
    \caption{Comparison between the HA and the EA lowest ten energies for the spiked oscillator with $\omega=1$, $a=0.5$ and various values of $\ell$ as indicated.}
    \label{tab:spiked_EAvsHA}
\resizebox{\linewidth}{!}{%
   \begin{tabular}{|c|c|c|c|c|c|c|c|} 
      \hline
      \multicolumn{2}{|c|}{$\ell=3$} & \multicolumn{2}{c|}{$\ell=4$}& \multicolumn{2}{c|}{$\ell=5$}&\multicolumn{2}{c|}{$\ell=6$}\\ % <-- Combining two cells with alignment c| and content 12.
      \hline
      \textbf{EA} & \textbf{HA} & \textbf{EA} & \textbf{HA}&\textbf{EA} & \textbf{HA}&\textbf{EA} & \textbf{HA}\\
      \hline
4.513881042&4.514506437&5.507596458&5.508494706&6.504630756&6.505852852&7.502959599&7.504555983\\
6.519685633&6.521143478&7.510109120&7.511986392&8.505796650&8.508146175&9.503462309&9.506335250\\
8.525296615&8.527889392&9.512506255&9.515663690&10.50685879&10.51063631&11.50386382&11.50831359\\
10.53071880&10.53474888&11.51478813&11.51952682&12.50781721&12.51332324&13.50416415&13.51049099\\
12.53595648&12.54172614&13.51695500&13.52357601&14.50867193&14.51620696&15.50436329&15.51286741\\
14.54101354&14.54882496&15.51900712&15.52781144&16.50942297&16.51928747&17.50446124&17.51544282\\
16.54589352&16.55604878&17.52094472&17.53223334&18.51007037&18.52256476&19.50445801&19.51821721\\
18.55059965&18.56340074&19.52276827&19.53684187&20.51061414&20.52603882&21.50435372&21.52119054\\
20.55513491&20.57088374&21.52446556&21.54163723&22.51105431&22.52970964&23.50414264&23.52436279\\
22.55950206&22.57850044&23.52659193&23.54661959&24.51139090&24.53357721&25.50408327&25.52773392\\
      \hline
    \end{tabular}
}%
  \end{center}
\end{table}

\subsection{The power-law potential with sextic singularity}
To show the power of our method, we choose a problem with sextic singularity,
\begin{equation}
\label{eqn:6ticV}
	    V(r)=\frac{\ell(\ell+1)}{2r^2}-\frac{b^2}{r^4}+\frac{1}{2}\frac{a^4}{r^6}
\end{equation}
where the real parameters $\{a,b\}$ are chosen such that $|b|>|a|$. By starting with the exponential Ansatz, we find $\mathcal{V}_0(\lambda,r)$ by replacing ${1}/{r^2}$, ${1}/{r^4}$, and ${1}/{r^6}$ in $V(r)$ with $v_2$, $v_4$, and $v_6$ from Appendix \ref{App:Ansatz}, respectively. Following the analysis in \ref{sec:PT_AIM}, we choose a coordinate transformation $y=1-2e^{-\lambda r}$ and $g(y)=(1-y)^3(1+y)^4$ to obtain the AIM second-order differential equation that matches the form of  Eq. \eqref{eqn:AIMeq}. As an illustration, we took $a=2$, $b=7$, considered several values of the angular momentum $\ell: 1- 6$ and calculated the corresponding energy spectrum as reported in Table \ref{tab:6tic_EA}. To examine the efficiency of our approach, we compared the calculated eigenvalues with those obtained via the higher-order finite difference scheme (HOFD) \cite{HOFD} as shown in Table \ref{tab:6tic_EA}. In our AIM computations, we took $y_0=2/15$ and $\lambda=0.5$. A good agreement between our results and the data from the HOFD has been observed. On the other hand, using the HA potential, we make a change of variable $y=2\tanh^2(\lambda r)-1$ and $g(y)=(1-y)^{10}(1+y)^{10}$. Solving Eq. \ref{eqn:termination} for $E^{(0)}$ with the use of $y_0=0$ and $\lambda=0.5$, we obtain the corresponding approximate energy spectrum of Eq. \ref{eqn:6ticV} as shown in Table. \ref{tab:6tic_HA}. The accuracy of those approximate ansatz is also seen from their plots versus the exact potential as illustrated in Fig. \ref{fig:sextic}.       

%Our choice of $\mathcal{V}_0(r)$ is 
%\begin{equation}
%    \mathcal{V}_0(r)=\frac{A}{{{\left( {{e}^{\lambda r}}-1 \right)}^{6}}}+\frac{B}{{{\left( {{e}^{\lambda r}}-1 \right)}^{5}}}+\frac{C}{{{\left( {{e}^{\lambda r}}-1 \right)}^{4}}}+\frac{D}{{{\left( {{e}^{\lambda r}}-1 \right)}^{3}}}+\frac{H}{{{\left( {{e}^{\lambda r}}-1 \right)}^{2}}}+\frac{F}{\left( {{e}^{\lambda r}}-1 \right)}+G
%\end{equation}
%where $A=\frac{{{\lambda }^{6}}{{a}^{4}}}{2}$, $B=\frac{3{{\lambda }^{6}}{{a}^{4}}}{2}$, $C=\frac{13{{\lambda }^{6}}{{a}^{4}}}{8}-{{\lambda }^{4}}{{b}^{2}}$, $ D=\frac{3{{\lambda }^{6}}{{a}^{4}}}{4}-2{{\lambda }^{4}}{{b}^{2}}$, and  
%\begin{align}
% & H=\frac{{{\lambda }^{2}}\ell (\ell +1)}{2}+\frac{31{{\lambda }^{6}}{{a}^{4}}}{240}-\frac{7}{6}{{\lambda }^{4}}{{b}^{2}} \\ 
% & F=A-B+C-D+H \\ 
% & G=\frac{F}{2}-\frac{5H}{12}+\frac{3D}{8}-\frac{251C}{720}+\frac{95B}{288}-\frac{19087A}{60480} \\ 
%\end{align}
%It is straightforward to show that $U(r)=V(r)-\mathcal{V}_0(r)$ vanishes at the origin. 

\begin{table}[h!]
  \begin{center}
    \caption{The energy spectrum of (\ref{eqn:6ticV}) for the parameters $a=2$, $b=7$, and different values of $\ell$ using the AIM with the EA potential versus the results obtain via the HOFD. For the AIM, we took 400 iterations.}
    \label{tab:6tic_EA}
\resizebox{\linewidth}{!}{%
   \begin{tabular}{|c|c|c|c|c|c|} 
      \hline
      \multicolumn{2}{|c|}{$\ell=0$} & \multicolumn{2}{c|}{$\ell=1$}& \multicolumn{2}{c|}{$\ell=2$}\\ % <-- Combining two cells with alignment c| and content 12.
      \hline
      \textbf{HOFD} & \textbf{AIM} & \textbf{HOFD} & \textbf{AIM}&\textbf{HOFD} & \textbf{AIM}\\
      \hline
      -198.053811725&-198.0537328&-194.485061786   &-194.4832736 &-187.382284129&-187.3817611\\
      -95.107391509&-95.10726998 &-92.507287345   &-92.50381690 &-87.351046558&-87.35019401\\
      -37.503430893&-37.50323491&-35.790300903 &-35.78702065 &-32.419721196&-32.41860615\\
      -10.649435153&-10.64906542 &-9.698773354&-9.696562310 &-7.867299662&-7.865513250 \\
      -1.563859663&-1.562902864 &-1.193393381&-1.190909350 &-0.540802068&-0.5360973375\\
     % -0.025300267&{\color{red}xxx} &   & && \\
        \hline
      \multicolumn{2}{|c|}{$\ell=3$}&
      \multicolumn{2}{c|}{$\ell=4$}&
      \multicolumn{2}{c|}{$\ell=5$}\\
      \hline
      \textbf{HOFD} & \textbf{AIM}&\textbf{HOFD} & \textbf{AIM}&\textbf{HOFD} & \textbf{AIM}\\
      \hline
       -176.816074818&-176.8143073 &-162.895376078&-162.8935434 &-145.771664441&-145.7696640 \\
      -79.728921145&-79.72578334 &-69.782624776&-69.77953504 &-57.714757854&-57.71149979 \\
      -27.507836187&-27.50448038 &-21.242775651&-21.23900784 &-13.908937252&-13.90422352 \\
     -5.305987630&-5.302008650 &-2.281242651&-2.274819916 &&\\
      \hline
    \end{tabular}
}%
  \end{center}
\end{table}

\begin{table}[h!]
  \begin{center}
    \caption{$a=2$, $b=7$ Hyperbolic Ansatz (HA) vs Exponential Ansatz (EA). In EA, we used $x_0=2/15$, $\lambda=0.5$, while in the HA, we took $x_0=0$, $\lambda=0.5$}
    \label{tab:6tic_HA}
\resizebox{\linewidth}{!}{%
   \begin{tabular}{|c|c|c|c|c|c|} 
      \hline
      \multicolumn{2}{|c|}{$\ell=0$} & \multicolumn{2}{c|}{$\ell=1$}& \multicolumn{2}{c|}{$\ell=2$}\\ % <-- Combining two cells with alignment c| and content 12.
      \hline
       \textbf{HA} & \textbf{EA} & \textbf{HA} & \textbf{EA}&\textbf{HA} & \textbf{EA}\\
      \hline
      -198.0491077&-198.0537328& -194.4740844  &-194.4832736 &-187.3713222&-187.3817611\\
      -95.10036570&-95.10726998 &-92.48865866 &-92.50381690&-87.33340002&-87.35019401\\
      -37.49199053&-37.50323491&-35.76710592  &-35.78702065 &-32.39420310&-32.41860615\\
      -10.62838677 &-10.64906542 &-9.667043170 &-9.696562310&-7.823876188&-7.865513250 \\
      -1.518415325 &-1.562902864 &-1.129380489 &-1.190909350&-0.4330104614 &-0.5360973375\\
      %{\color{red}xxx}&{\color{red}xxx}&&&&\\
      %--&--&&&&\\
        \hline
      \multicolumn{2}{|c|}{$\ell=3$}&
      \multicolumn{2}{c|}{$\ell=4$}&
      \multicolumn{2}{c|}{$\ell=5$}\\
      \hline
       \textbf{HA} & \textbf{EA} & \textbf{HA} & \textbf{EA}&\textbf{HA} & \textbf{EA}\\
      \hline
       -176.8021730&-176.8143073 &-162.8781685&-162.8935434 &-145.7473592&-145.7696640 \\
     -79.70705001&-79.72578334 &-69.75589780&-69.77953504 &-57.67625301&-57.71149979 \\
      -27.47327862&-27.50448038 &-21.19529103&-21.23900784 &-13.83791731&-13.90422352 \\
     -5.239053277&-5.302008650 &-2.172355430&-2.274819916 &&\\
      \hline
    \end{tabular}
}%
  \end{center}
\end{table}

\section{Conclusion}
Based on the zeroth order perturbation theory, we have proposed, in the AIM framework, two approximate schemes of the radial Schr\"odinger equation involving power-law potentials. We provided several examples to illustrate the efficiency of our approach. It was possible to study the challenging case of the supersingular sextic potential despite the failure of the traditional AIM due to the high instability in finding the root of the termination condition. Since we use a zeroth order perturbation method, our approach has a minimal computational cost compared to other techniques used to study such classes of singular potentials. The present work opens a door to study various eigenvalues in the AIM framework that were previously difficult to treat. We may emphasize that we have indirectly solved new classes of eigenvalue problems generated by the potential Ansatz and can be treated using AIM.

\section*{Acknowledgments} Partial financial support of this work under Grant No. GP249507 from the
Natural Sciences and Engineering Research Council of Canada
 is gratefully acknowledged.

\appendix
%=========================
\section{Potential Ansatz up to sextic singularities}
\label{App:Ansatz} 
\subsection{The Hulthen type Ansatz}
Let $u_n=\frac{1}{r^n}$ and the potential we are willing to study $V(r)$ is a linear combination of such terms. The Ansatz potential $\mathcal{V}_0(r)$ is obtained by replacing every $u_n$ with an alternative expansion $v_n$. Here, we list the first six expansions 
\begin{align*}
    \frac{1}{\lambda}v_1&=\frac{1}{{{e}^{\lambda r}}-1}+\frac{1}{2},\\
    \frac{1}{{{\lambda }^{2}}}v_2&={{\left( \frac{1}{{{e}^{\lambda r}}-1} \right)}^{2}}+\frac{1}{\lambda r}-\frac{5}{12},\\
    \frac{1}{{{\lambda }^{3}}}v_3&={{\left( \frac{1}{{{e}^{\lambda r}}-1} \right)}^{3}}+\frac{3}{2}{{\left( \frac{1}{{{e}^{\lambda r}}-1} \right)}^{2}}+\frac{1}{2}\frac{1}{{{e}^{\lambda r}}-1},\\
    \frac{1}{{{\lambda }^{4}}}v_4&={{\left( \frac{1}{{{e}^{\lambda r}}-1} \right)}^{4}}+2{{\left( \frac{1}{{{e}^{\lambda r}}-1} \right)}^{3}}+\frac{7}{6}{{\left( \frac{1}{{{e}^{\lambda r}}-1} \right)}^{2}}+\frac{1}{6}\frac{1}{{{e}^{\lambda x}}-1}-\frac{1}{720},\\
    \frac{1}{{{\lambda }^{5}}}v_5&={{\left( \frac{1}{{{e}^{\lambda r}}-1} \right)}^{5}}+\frac{5}{2}{{\left( \frac{1}{{{e}^{\lambda r}}-1} \right)}^{4}}+\frac{25}{12}{{\left( \frac{1}{{{e}^{\lambda r}}-1} \right)}^{3}}+\frac{15}{24}{{\left( \frac{1}{{{e}^{\lambda r}}-1} \right)}^{2}}+\frac{1}{24}\frac{1}{{{e}^{\lambda r}}-1},\\
    \frac{1}{{{\lambda }^{6}}}v_6&={{\left( \frac{1}{{{e}^{\lambda r}}-1} \right)}^{6}}+3{{\left( \frac{1}{{{e}^{\lambda r}}-1} \right)}^{5}}+\frac{13}{4}{{\left( \frac{1}{{{e}^{\lambda r}}-1} \right)}^{4}}+\frac{3}{2}{{\left( \frac{1}{{{e}^{\lambda r}}-1} \right)}^{3}}\\
    &\quad+\frac{31}{120}{{\left( \frac{1}{{{e}^{\lambda r}}-1} \right)}^{2}}+\frac{1}{120}\frac{1}{{{e}^{\lambda r}}-1}+\frac{1}{\text{30240}}.
\end{align*}

\subsection{The hyperbolic Ansatz} 
Similar to the previous section, we obtain an alternative potential Ansatz by replacing every $u_n$ by $w_n$, where we list the first six expansions below 

\begin{align*}
    \frac{1}{\lambda }w_1&=\frac{1}{\sinh \left( \lambda r \right)}\\
    \frac{1}{{{\lambda }^{2}}}w_2&=\frac{1}{{{\sinh }^{2}}\left( \lambda r \right)}+\frac{1}{3},\\
    \frac{1}{{{\lambda }^{3}}}w_3&=\frac{1}{{{\sinh }^{3}}\left( \lambda r \right)}+\frac{1}{2}\frac{1}{\sinh \left( \lambda r \right)},\\
    \frac{1}{{{\lambda }^{4}}}w_4&=\frac{1}{{{\sinh }^{4}}\left( \lambda r \right)}+\frac{2}{3}\frac{1}{{{\sinh }^{2}}\left( \lambda r \right)}-\frac{1}{45},\\
    \frac{1}{{{\lambda }^{5}}}w_5&=\frac{1}{{{\sinh }^{5}}\left( \lambda r \right)}+\frac{5}{6}\frac{1}{{{\sinh }^{3}}\left( \lambda r \right)}+\frac{5}{12}\frac{1}{\sinh \left( \lambda r \right)}-\frac{3}{8}\frac{1}{\sinh \left( \lambda r \right)},\\
    \frac{1}{{{\lambda }^{6}}}w_6&=\frac{1}{{{\sinh }^{6}}\left( \lambda r \right)}+\frac{1}{{{\sinh }^{4}}\left( \lambda r \right)}+\frac{2}{15}\frac{1}{{{\sinh }^{2}}\left( \lambda r \right)}+\frac{2}{945}.
\end{align*}

%%%End of the excluded block

%


\begin{thebibliography}{}
\bibitem{d2000}Shi-Hai dong, \emph{Exact Solutions of the Two-Dimensional Schrödinger Equation with Certain Central Potentials}. International Journal of Theoretical Physics 39, 1119 - 1128 (2000).
\bibitem{a2002} A. D. Alhaidari, \emph{
Solutions of the nonrelativistic wave equation with position-dependent effective mass}.
Phys. Rev. A 66, 042116 (2002).
\bibitem{j2019} G. Junker \emph{Supersymmetric Methods in Quantum, Statistical and Solid State Physics}. IOP Publishing (2019).
\bibitem{s2013} J. Sesma, \emph{Exact solution of the Schrödinger equation with a Lennard–Jones potential}. J Math Chem 51, 1881 - 1896 (2013). https://doi.org/10.1007/s10910-013-0189-9.
\bibitem{d2015} Daniel Baye, \emph{The Lagrange-mesh method}. Physics Reports Vol. 565, 1 -107 (2015).
\bibitem{HOFD} A.D. Alhaidari, I.A. Assi, and A. Mebirouk. \emph{Bound states of a quartic and sextic inverse-power-law potential for all angular momenta}. The European Physical Journal Plus, 136(4), 1-12 (2021).
\bibitem{FD2} A. Mebirouk, S. Bouheroum-Mentri and L. Aceto, Approximation of eigenvalues of 
Sturm-Liouville problems defined on a semi-infinite domain, Appl. Math. Comput. 369 124823 (2020).
\bibitem{AIM1} H. Ciftci, R. L. Hall, and N. Saad, J. Phys. A: Math. \& Gen., \textbf{36} (47), 11807 (2003).

\bibitem{AIM_SP1} B. Champion, R.L. Hall, and N. Saad. \emph{Asymptotic iteration method for singular potentials}. International Journal of Modern Physics A, 23(09), 1405-1415 (2008).
\bibitem{AIM_SP2} T. Barakat. \emph{The asymptotic iteration method for the eigenenergies of the Schrödinger equation with the potential $V(r)=-Z/r+g\,r+\lambda r^2$}. Journal of Physics A: Mathematical and General, 39(4), 823 (2006).

\bibitem{AIM_SP3} P. Amore and F.M. Fernández,\emph{Comment on an application of the asymptotic iteration method to a perturbed Coulomb model}. Journal of Physics A: Mathematical and General, 39(33), 10491 (2006).

\bibitem{AIM_PT}
H.I. Alrebdi and T. Barakat. \emph{Optimal perturbation technique within the Asymptotic Iteration Method for heavy-light meson mass splittings.} Universe, 7(6), 180 (2021).

\bibitem{FF} F. M. Fernandez. \emph{Introduction to Perturbation Theory in Quantum Mechanics} (1st ed.). CRC Press (2000). https://doi.org/10.1201/9781420039641



\bibitem{Spiked1} R. L. Hall, N. Saad, and A. B. von Keviczky, \emph{Generalized spiked harmonic oscillator}. J. 
Phys. A 34, 1169 (2001).
\bibitem{Spiked2} R. L. Hall, N. Saad, and A. B. Keviczky, \emph{Spiked harmonic oscillators}, J. Math. Phys. 43, 94 (2002).
\bibitem{Spiked3} F. J. G\'{o}mez and J. Sesma, \emph{Spiked oscillators: exact solution}. J. Phys. A 43, 385302 (2010).
\bibitem{Spiked4} A. K. Roy, \emph{Ground and excited states of spherically symmetric potentials through an 
imaginary-time evolution method: application to spiked harmonic oscillators}. J. Math. 
Chem. 52, 2645 (2014).
\bibitem{Spiked5} A. K. Roy and A. F. Jalbout, \emph{Bound states of the generalized spiked harmonic oscillator,}
J. Molec. Struct. 853, 27 (2008).
\bibitem{Spiked6} O. Mustafa and M. Odeh, \emph{Bound states for spiked harmonic oscillators and truncated 
Coulomb potentials}, J. Phys. A 32, 3055 (1999).
\bibitem{Spiked7} A.D. Alhaidari. \emph{Finite series representation for the bound states of a spiked isotropic oscillator with inverse-quartic singularity}. Modern Physics Letters A, 37(07), 2250045 (2022).

\end{thebibliography}
\end{document}